\documentclass{article}
\usepackage{spconf,amsmath,graphicx}
\usepackage{tabularx}
\usepackage[ruled,linesnumbered]{algorithm2e}
\usepackage[labelfont=bf]{caption} 
\usepackage{hyperref,url}
\usepackage{multirow}
\usepackage{booktabs}
\usepackage{cite} 
\usepackage{amssymb} 
\usepackage{cleveref}

\ninept

\title{SkiM: Skipping Memory LSTM for Low-Latency Real-Time \\
Continuous Speech Separation}
\name{Chenda Li$^1$, Lei Yang$^2$, Weiqin Wang$^2$, Yanmin Qian$^1$, \thanks{
Yanmin Qian is the corresponding author.}}
\address{
$^1$MoE Key Lab of Artificial Intelligence, AI Institute \\
X-LANCE Lab, Department of Computer Science and Engineering \\
Shanghai Jiao Tong University, Shanghai, China \\
$^2$Samsung Research China - Beijing (SRC-B)\\ 
    $^1$\{lichenda1996,yanminqian\}@sjtu.edu.cn,
    $^2$\{lei81.yang,wq88.wang\}@samsung.com
    }
\begin{document}
%
\maketitle
\begin{abstract}
Continuous speech separation for meeting pre-processing has recently become a focused research topic.
Compared to the data in utterance-level speech separation, the meeting-style audio stream lasts longer, has an uncertain number of speakers.
We adopt the time-domain speech separation method and the recently proposed Graph-PIT to build a super low-latency online speech separation model, which is very important for the real application. 
The low-latency time-domain encoder with a small stride leads to an extremely long feature sequence.
We proposed a simple yet efficient model named Skipping Memory (SkiM) for the long sequence modeling.
Experimental results show that SkiM achieves on par or even better separation performance than DPRNN.
Meanwhile, the computational cost of SkiM is reduced by $75\%$ compared to DPRNN.
The strong long sequence modeling capability and low computational cost make SkiM a suitable model for online CSS applications.
Our fastest real-time model gets $17.1$ dB signal-to-distortion (SDR) improvement with less than $1$-millisecond latency in the simulated meeting-style evaluation.

\end{abstract}


\section{Introduction}
\label{sec:intro}

Following the Deep Clustering \cite{hershey2016deep} and permutation invariant training (PIT) \cite{yu2017permutation,kolbaek2017multitalker}, the speech separation driven by neural network has been rapidly developed \cite{chen2017deep,wang2018end,luo2018tasnet,delcroix2018single,luo2019conv,wang2019voicefilter,luo2020dual,subakanAttentionAllYou2021}.
Current state of the art systems \cite{luo2020dual,subakanAttentionAllYou2021} show impressive performance on the utterance-level benchmark \cite{hershey2016deep}.
However, how to effectively bring the speech separation systems into the real application (e.g., meeting processing) remains a challenge.
Compared to the utterance-level benchmark like WSJ0-2mix \cite{hershey2016deep}, the meeting-style data involves more speakers, lasts longer, and contains a large amount of non-overlapped or silence clips.
But most conventional speech separation systems are trained with a fixed number of speakers on short well-segmented fully-overlapped data.

Recent researches attempt to extend the utterance-level speech separation to the Continuous Speech Separation (CSS) \cite{yoshioka2019css,chen2020continuous}.
One of the most straightforward extension for CSS consists of three steps: \textit{Segmentation}, \textit{Separation} and \textit{Stitching} (3S).
It firstly segments the long recording into smaller windows. 
When the window size is small enough, it is reasonable to assume each window at most involves $2$ (or $3$) speakers.
Secondly, conventional utterance-level speech separation can be applied for each window.
In the last step named \textit{stitching}, the permutation order of separated overlap-free audios from windows is aligned. 
After stitching, the CSS system can generate continuous overlap-free speech.
The 3S paradigm has been adopted by most of the recent CSS works \cite{wangMultimicrophoneComplexSpectral2020,li2021dualpath,chenDonShootButterfly2021,liDualPathModelingLong2021,chenContinuousSpeechSeparation2021,chenUltraFastSpeech2021,hanContinuousSpeechSeparation}.
However, there are some shortcomings in this 3S paradigm.
First, the processing of each window usually does not depend on the others, which may lead to sub-optimal performance.
Second, the stitching stage compares the similarity of an overlap region between the adjacent separated windows.
Therefore, the computational overhead is introduced for separating the overlap region twice. 
Last, the choice of the window length is a dilemma. The small window size may hurt the separation performance and stitching stability, and large window size may break the assumption of maximum speaker number in one window.
A recent research \cite{neumann21_interspeech} proposed Graph-PIT to train the separation system with the entire meeting.
Without the segmentation, the above issues of 3S CSS are avoided.


In this paper, we explore the approach for building a low-latency real-time CSS system. 
To achieve the target of super-low ideal latency, we adopt the time-domain separation model with a small encoder stride.
Combining with the Graph-PIT meeting-level training, we need to handle the extremely long feature sequence.
For example, in our causal system of $0.6$ millisecond ideal latency, the separator needs to process about $50$k-frame feature sequence for a $30$-second meeting clip.
This task poses a big challenge to conventional sequence models like recurrent neural network (RNN) or Transformer \cite{vaswani2017attention}.
The previous proposed dual-path RNN (DPRNN) \cite{luo2020dual} is a qualified candidate.
But its alternately inter- and intra-chunk modeling strategy requires a high computational cost, which is unfriendly to the low-power devices in the real-time application.

We propose a simple extension to the long short-term memory (LSTM) \cite{hochreiter1997long}, which is named skipping memory (SkiM).
The idea behind SkiM is inspired by DPRNN \cite{luo2020dual}, i.e., alternately modeling for the local and global information.
DPRNN uses an inter-chunk RNN to model the long-span feature frame-by-frame, which might be too fine-grained.
SkiM abandons the inter-chunk RNN and uses a more efficient way to share the global-aware hidden and cell states between the local LSTMs.
For the long-span information modeling, the SkiM model just \textbf{skims} the long sequence rather than peruse it. 
Thus the computational cost can be hugely reduced.
Our experiments show that the computation cost of SkiM can be reduced by $75\%$ compared with DPRNN.
Meanwhile, in the CSS task, the separation performance of SkiM is on par or even better than DPRNN.
The low computational dependence makes SkiM a suitable model for low-latency real-time CSS.

\begin{figure*}[t]

  \centering
\includegraphics[width=0.95\linewidth]{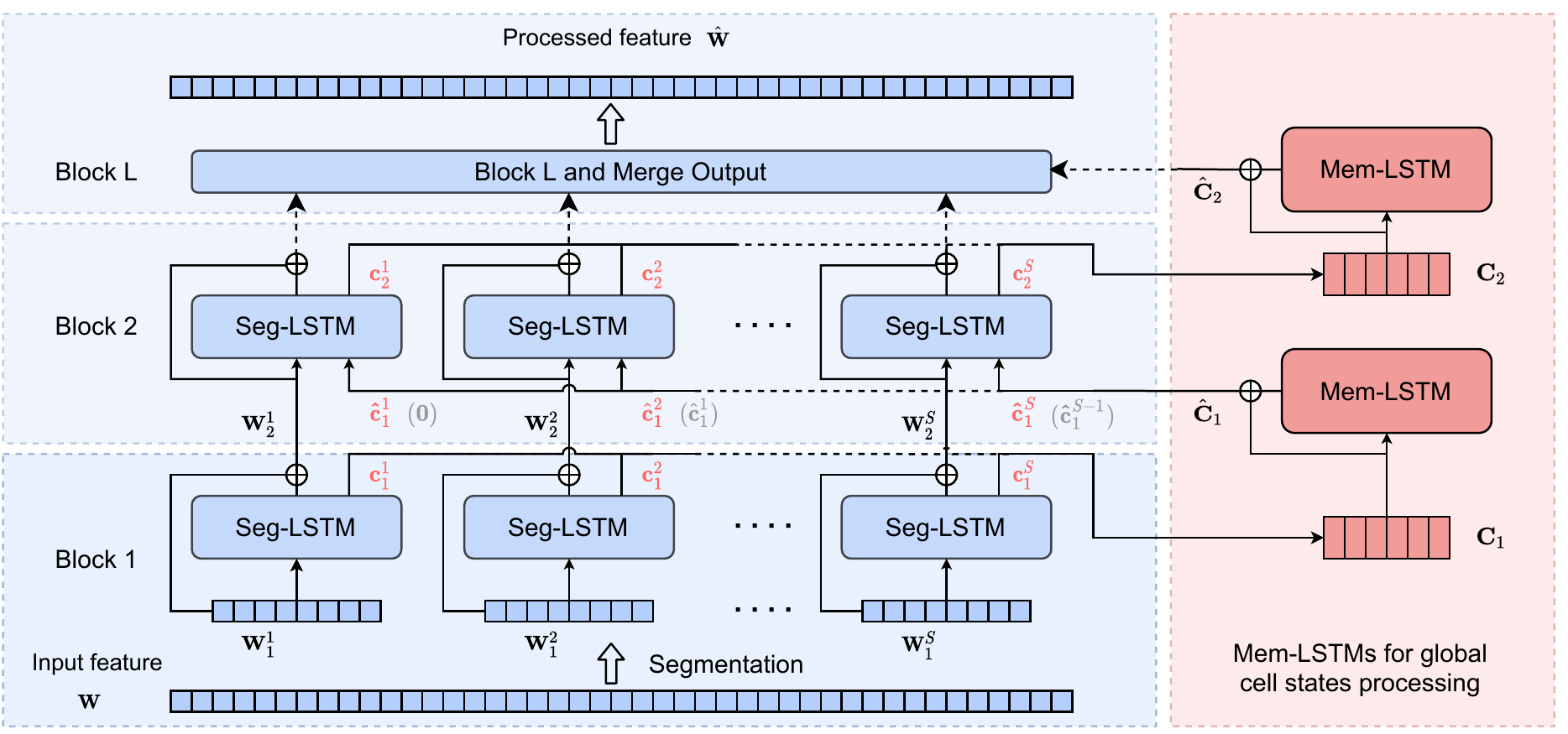}

\caption{The proposed Skipping Memory Net. 
Seg-LSTMs (blue) process small segments for local modeling. 
Each segment generates a cell state $\mathbf{c}^{s}$.
Mem-LSTMs (red) process the cell states to sync global information between different segments. 
The processed cell states are then used to initialize Seg-LSTM in the next SkiM block. 
The gray font $\hat{\mathbf{c}}$ is for causal system.
The processing flow for hidden state $\mathbf{h}$ are the same as that for the cell state $\mathbf{c}$; Memory LSTMs for the hidden state is omitted in this figure for clarity. }
\label{fig:skim}
\end{figure*}

\section{SkiM for online continuous speech separation}

\subsection{Backbone}

Our model follows the backbone of the TasNet \cite{luo2018tasnet,luo2019conv}, which consists of an encoder, a separator and a decoder. 
Denote the input time-domain speech mixture as $\mathbf{x} \in \mathbb{R}^{L \times C} $, where $L$ is the length of the signal and $C$ is the number of channels \footnote{In this paper, $C$ is $1$ since we only discuss the single-channel input.}.
The encoder contains a single layer 1-D convolution, and it maps the $\mathbf{x}$ into the high-dimensional input feature $\mathbf{W}\in \mathbb{R}^{T \times N}$ of length $T$ and dimension $N$. 
The separator is a sequence model. It processes the input feature and generates processed embedding sequences for each output channel.
This section will introduce the proposed SkiM separator.
The decoder is made up of a 1-D transposed convolution layer, which transfers the processed embedding back to the time domain signal.

\subsection{Review of LSTM}
The proposed SkiM separator is based on long short-term memory (LSTM) \cite{hochreiter1997long}.
We first review the mapping function of a typical LSTM layer:
\begin{align}
	\hat{\mathbf{W}}, \hat{\mathbf{c}}, \hat{\mathbf{h}} = \operatorname{LSTM}(\mathbf{W}, \mathbf{c}, \mathbf{h})
\end{align}
where $\mathbf{W} \in \mathbb{R}^{T \times N} $is the input sequence.
$\mathbf{c}$ and $\mathbf{h}$ are the initial cell state vector and hidden state, respectively. 
$\mathbf{c}$ and $\mathbf{h}$ are usually initialized with $\mathbf{0}$ in most applications.
$\hat{\mathbf{W}} \in \mathbb{R}^{T \times N} $ is the output sequence.
$\hat{\mathbf{c}}$ is the update cell state, which is regarded as to encode the long-term memory of the whole sequence \cite{hochreiter1997long}.
$\hat{\mathbf{h}}$ is the updated hidden state as well as the last-step's output in $ \hat{\mathbf{h}} = \hat{\mathbf{W}}[T,:]$. 
In LSTM, $\hat{\mathbf{h}}$ is also considered as the short-term memory of the processed sequence.

\subsection{Skipping Memory Separator}
\label{sec:skim_separator}

Denote the input feature as $\mathbf{W} \in \mathbb{R}^{T \times N}$, where $T$ and $N$ are the number of time steps and feature dimension, respectively.
In this time-domain CSS task, $T$ is usually very large ($T > 1 \times 10^4$) .
The input feature $\mathbf{W}$ are split into smaller segments $\left[\mathbf{W}_{1}^{1}, \mathbf{W}_{1}^{2}, \cdots,  \mathbf{W}_{1}^{S} \right]$, where $\mathbf{W}_{1}^{s} = \mathbf{W}[sK-K:sK,:] \in \mathbb{R}^{K \times N} $ , $s={1, \dots, S}$, $S$ is the number of segments, and $K$ is the segment length.

As Fig. \ref{fig:skim} shows, SkiM is made up of $L$ basic blocks, and each block contains a Seg-LSTM layer.
Seg-LSTM is used to process a small segment $\mathbf{W}^{s}$.
Denote $\left[\mathbf{W}_{l}^{1}, \mathbf{W}_{l}^{2}, \cdots,  \mathbf{W}_{l}^{S} \right]$ as the input of the $l$-th SkiM block, and the mapping function of Seg-LSTM can be formulated as: 

\begin{align}
	\label{eq:seg-lstm}
	\bar{\mathbf{W}}_{l + 1}^{s}, \mathbf{c}_{l + 1}^{s}, \mathbf{h}_{l + 1}^{s} &= \operatorname{Seg-LSTM}\left(\mathbf{W}_{l}^{s}, \hat{\mathbf{c}}_{l}^{s}, \hat{\mathbf{h}}_{l}^{s}\right) \\
	\mathbf{W}_{l + 1}^{s} &= \operatorname{LN}(\bar{\mathbf{W}}_{l + 1}^{s}) + \mathbf{W}_{l}^{s} 
\end{align}
where $l = 1, \dots, L$, $\hat{\mathbf{c}}_{1}^{s} = \mathbf{0}$, $\hat{\mathbf{h}}_{1}^{s} = \mathbf{0}$, and $\operatorname{LN}$ is the layer normalization operation in the residual connection \cite{he2016identity}.
The processed cell state $\mathbf{c}_{l+1}^{s}$ and hidden state $\mathbf{c}_{l+1}^{s}$ from each individual segment $\mathbf{W}^{s}$ are collected as $\mathbf{C}_{l+1} = [\mathbf{c}_{l+1}^{1}, \dots, \mathbf{c}_{l+1}^{S}]$ and $\mathbf{H}_{l+1} = [\mathbf{h}_{l+1}^{1}, \dots, \mathbf{h}_{l+1}^{S}]$.
$\mathbf{C}_{l+1}$ and $\mathbf{H}_{l+1}$ are then fed into another Mem-LSTM for cross-segment processing.
\begin{align}
	\bar{\mathbf{C}}_{l+1} &= \operatorname{Mem-LSTM}_{c}(\mathbf{C}_{l+1}), \\
	\bar{\mathbf{H}}_{l+1} &= \operatorname{Mem-LSTM}_{h}(\mathbf{H}_{l+1}), \\
	\hat{\mathbf{C}}_{l+1} &= \operatorname{LN}({\bar{\mathbf{C}}_{l+1}}) + \mathbf{C}_{l+1} \\
	\hat{\mathbf{H}}_{l+1} &= \operatorname{LN}({\bar{\mathbf{H}}_{l+1}}) + \mathbf{H}_{l+1}
\end{align}

The global-synced memory states $\hat{\mathbf{C}}_{l+1} = [\hat{\mathbf{c}}_{l+1}^{1}, \dots, \hat{\mathbf{c}}_{l+1}^{S}]$ and $\hat{\mathbf{H}}_{l+1} = [\hat{\mathbf{h}}_{l+1}^{1}, \dots, \hat{\mathbf{h}}_{l+1}^{S}]$  are used as the initial states of the Seg-LSTM in the next block.
The last SkiM block's output $\left[ \mathbf{W}_{L + 1}^{1}, \cdots, \mathbf{W}_{L + 1}^{S} \right]$ is merged as $\hat{\mathbf{W}}$. $\hat{\mathbf{W}}$ is then sent into fully connect (FC) layers and transposed convolution decoder to estimate the separated signal for each output channel.

\subsection{Causal SkiM for online implementation}
For offline processing, the LSTM layers in SkiM could be bi-directional to achieve the best performance.
While in the online processing, it requires SkiM to be a causal system.
Thus, we use uni-directional LSTMs in SkiM, and Eq. \ref{eq:seg-lstm} should be revised to:
\begin{align}
	\bar{\mathbf{W}}_{l + 1}^{s}, \mathbf{c}_{l + 1}^{s}, \mathbf{h}_{l + 1}^{s} &= \operatorname{Seg-LSTM}\left(\mathbf{W}_{l}^{s}, \hat{\mathbf{c}}_{l}^{s-1}, \hat{\mathbf{h}}_{l}^{s-1}\right) 
\end{align}
where $s = 1, \cdots, S$, $\hat{\mathbf{c}}_{l}^{0} = \mathbf{0}, \hat{\mathbf{h}}_{l}^{0}  = \mathbf{0}$. Then the processing of the $s$-th segment only depends on the segments $1, \cdots, s-1$.

\subsection{Discussion on Computational Optimization }

To achieve lower ideal latency for the online CSS system, one of the most practical way is to squeeze the stride size of the convolution encoder.
But that will also lead to a huge number of time steps $T$ for the continuous input feature.

A recently proposed dual-path RNN (DPRNN) \cite{luo2020dual} is an efficient model for long sequence modeling.
It uses an \textit{intra-chunk} and \textit{inter-chunk} RNN to capture the local information and long-span information, respectively.
However, the frame-by-frame modeling manner of the \textit{inter-chunk} RNN is too fine-grained, which may be computationally expensive for real-time processing.

Compared with the DPRNN of a similar model size, SkiM greatly reduces the computation cost, and it is better for the application that requires real-time performance.
There are two main reasons for the reduction in the amount of calculation.
Firstly, SkiM uses Mem-LSTM for long-span modeling with skipping memory (hidden and cell states) manner.
Each segment produces a cell and hidden state vectors that encode the knowledge for the local segment, and Mem-LSTM only needs to process the state sequence.
Secondly, in the segmentation stage, we abandon the overlap region (usually $50\%$ in DPRNN) between the adjacent segments to reduce the computation cost.

\section{Experiments}

\subsection{Dataset}
We use a simulated meeting style dataset derived from LibriSpeech \cite{panayotov2015librispeech}.
The speech is noisy and reverberant.
Each audio sample in the dataset is a simulated meeting session lasting for about $90$ seconds, contains $3$-$5$ active speakers, and the overlap ratio of utterances in the session is between $50\%$ and $80\%$.
The simulated dataset is also used in our previous work \cite{liDualPathModelingLong2021}, readers can refer to it for a detailed description.
In the training stage, we randomly clip the $90$-seconds audio into a fixed length of $30$ seconds for saving the GPU memory.
While in the inference stage, we input the entire meeting session to the model.
The sampling rate of the audios is $16$ kHz.

\subsection{Training Criterion}

A recently proposed session-level training criterion named Graph-PIT \cite{neumann21_interspeech} is adopted in this work.
To prepare the training label, $P$ reference utterances in the meeting need to be put into the $Q$ output channels. $P$ is usually much bigger than $Q$, and Graph-PIT solves the label permutation as a graph coloring problem.

Graph-PIT allows us to train the model with a meeting-level PIT manner, i.e., the loss computation is between the entire output signals and the overlap-free target signals.
The thresholded signal-to-distortion (tSDR) loss \cite{wisdom2020unsupervised} is used together with the Graph-PIT.
The $\text{SNR}_{max}$ and $\varepsilon$ of tSDR are set to $20$dB and $10^{-6}$, respectively.

\subsection{Model Configurations}

We perform experiments for both of the causal and non-causal SkiM systems.
All the SkiM models consist of $4$ SkiM blocks.
In each SkiM block, the LSTMs's hidden dimension is $256$.
The LSTMs in the causal systems are bi-directional, thus they have double parameters.
The segment size $S$, i.e., the length of feature processed by Seg-LSTM, is set to $150$.
The layer normalization operation in the non-causal SkiM is global, while it is only performed on the feature dimension in the causal SkiM.
For comparison, we also implement temporal convolutional network (TCN) \cite{luo2019conv} and DPRNN in this task.
The DPRNN has $4$ DPRNN blocks, and it has the same width as the SkiM model.

The models are trained with ESPNet-SE toolkit \cite{li2021espnet}.
The Adam \cite{kingma2014adam} optimizer is used for all the models in training.
The initial learning rate is set to $10^{-3}$, and it is reduced by a factor of $0.97$ for every epoch, and all models are trained for $100$ epochs.
The $L_2$-norm of the gradient is clipped to $5$ in model optimization.

\subsection{Evaluation}

We compute the model on the simulated testing set with different evaluations.
The first one is the best SDR improvement computed with Graph-PIT between the estimated signals and the overlap-free labels.
The second one is the Graph-PIT optimized Short Term Objective Intelligibility (STOI) \cite{taal2010short}.
The optimized STOI and SDR evaluates the overall quality of the processed meeting sessions.
Since the meeting-style data contains plenty of the single-talker and silent regions, the overall score may not show the true separation performance.
To figure out the separation performance on the harder case, we split the reference signals and estimated signals into small windows of $2$ seconds and only evaluate the SDR improvement for windows that have more than $50\%$ speech overlap.

\subsection{Results on Simulated Meeting}

\begin{table}[]
\caption{The overall STOI, SDR improvement (SDRi) and high-overlap SDR improvement (SDRi50) comparison on different models.}
\label{tab:skim_main}
\centering
    \setlength{\tabcolsep}{1.0mm}{
\scalebox{0.9}{
\begin{tabular}{c|cccccccc}
\toprule
Model      & Causal  & \multicolumn{1}{c}{\begin{tabular}[c]{@{}c@{}}Stride\\Size\end{tabular}}  &  \begin{tabular}[c]{@{}c@{}} Model\\size (M)\end{tabular} & \multicolumn{1}{c}{\begin{tabular}[c]{@{}c@{}}MACs\\ (G/s)\end{tabular}} & STOI   &\multicolumn{1}{c}{\begin{tabular}[c]{@{}c@{}}SDRi\\ (dB)\end{tabular}}  & \multicolumn{1}{c}{\begin{tabular}[c]{@{}c@{}}SDRi50\\ (dB)\end{tabular}}  \\ 
\midrule
TCN & no &  20 & 3.4 & 2.7 & 0.732 & 13.5 & 5.9 \\
\midrule
\multirow{3}{*}{DPRNN} & no & 20 & 9.6 & 14.6 & 0.767 & \textbf{19.2} & 9.0 \\
 & yes & 20 & 4.9 & 7.5 & 0.738 & 16.6 & 7.6 \\
 & yes & 10 & 4.9 & 14.7 & 0.737 & 16.8 & 7.7 \\
 \midrule
 \multirow{3}{*}{SkiM} & no & 20 & 15.9 & \textbf{3.8} & \textbf{0.768} & 18.7 & \textbf{9.2} \\
 & yes & 20 & 6.0 & \textbf{2.0} & \textbf{0.749} & \textbf{17.3} & \textbf{8.0} \\
 & yes & 10 & 6.0 & \textbf{3.9} & \textbf{0.745} & \textbf{17.1} & \textbf{7.8} \\
 
 \bottomrule
\end{tabular}
}
}
\end{table}

Table.\ref{tab:skim_main} listed the experimental results of different models.
The stride size of the convolution encoders is listed in the table, which is half of the convolution kernel size.
The ideal latency of the causal systems can be inferred from the stride size.
For example, for the $16$kHz data, the system of stride size $10$ has the ideal latency of about $0.6$ millisecond.
We also evaluated the computational cost with the number of multiplier–accumulator operations (MACs) per second, which is tested out on a $30$ seconds input audio.

From the results in table.\ref{tab:skim_main}, we find that TCN gets the worst performance.
The reason should be that the receptive field of TCN is much smaller than the sequence length in this CSS task;
The meeting-level Graph-PIT that we used may also require the model's global awareness for better optimization.
DPRNN and SkiM are designed for the long sequence modeling in CSS, and experiments confirm that they are much better than the TCN model. 

Doing comparison between the SkiM and DPRNN, we can see that SkiM gets on par or better separation performance for both causal and non-causal systems.
More importantly, the computational cost of the SkiM model is reduced by $\sim 75\%$ compared to DPRNN.
Benefits from less computation cost, the newly proposed SkiM is a better candidate for the real-time CSS that is deployed on low-power devices.

\subsection{Real-time Evaluation}
To verify the feasibility of deploying a real-time CSS on low-power devices, we test the real-time factor (RTF) and latency for the causal SkiM models.
We test the models on an old Intel CPU (8 years ago, 1.9GHz, with AVX instruction), and limit the Intel Math Kernel Library (iMKL) to run with a single thread. 
Now most of smartphones are more efficient than that Intel CPU.
Table.\ref{tab:skim_rtf} lists the RTFs and latency of the real-time CSS systems.
The RTFs of proposed SkiM models are all smaller than $1.0$, which means they have promising real-time processing capabilities on low-power devices.
Our fastest SkiM model achieves a low-latency of less than $1.0$ millisecond.

\begin{table}[h]
\caption{Real-time factor (RTF) and latency evaluation for causal models. Actual latency is listed for the models with RTF $< 1.0$.  }
\label{tab:skim_rtf}
\centering
    \setlength{\tabcolsep}{1.3mm}{
    \scalebox{0.9}{
\begin{tabular}{c|cc|ccc}
\toprule
Model & Stride size & Ideal latency &
\multicolumn{1}{c}{\begin{tabular}[c]{@{}c@{}}MACs\\ (G/s)\end{tabular}}
& RTF  & Latency\\
\midrule 
\multirow{2}{*}{DPRNN} & 20 & 1.25 ms & 7.5 & 0.98 & 2.47 ms\\
 & 10 & 0.625 ms & 14.7 & 1.98 & null \\
\midrule
\multirow{2}{*}{SkiM} & 20 & 1.25 ms & \textbf{2.0} & \textbf{0.23} & 1.54 ms\\
 & 10 & 0.625 ms & 3.9 & 0.46 & \textbf{0.92} ms \\
 
 \bottomrule
\end{tabular}
}
}
\end{table}

\begin{table}[]
\caption{Ablation studies for Mem-LSTMs in SkiM}
\label{tab:ablation}
\centering
\scalebox{0.9}{
\begin{tabular}{c|cccccc}
\toprule
Model      & \begin{tabular}[c]{@{}c@{}} Mem-\\LSTM\end{tabular}&  \begin{tabular}[c]{@{}c@{}} Model\\ size (M)\end{tabular} & \multicolumn{1}{c}{\begin{tabular}[c]{@{}c@{}}MACs\\ (G/s)\end{tabular}} & \multicolumn{1}{c}{\begin{tabular}[c]{@{}c@{}}SDRi\\ (dB)\end{tabular}}  & \multicolumn{1}{c}{\begin{tabular}[c]{@{}c@{}}SDRi50\\ (dB)\end{tabular}}  \\ 
\midrule
 \multirow{4}{*}{SkiM} & h, c & 15.9 & 3.8 & 18.7 & 9.2 \\
 & h, 0 & 10.4 & 3.8 & 17.8 & 8.6 \\
 & 0, c & 10.4 & 3.8 & 15.6 & 7.8 \\
 & 0, 0 & 4.9 & 3.8 & 12.5 & 7.1 \\
 & id, id & 4.9 & 3.8 & 12.5 & 7.1 \\
 \bottomrule
\end{tabular}
}
\end{table}

\subsection{Ablation Studies}

We perform the ablation studies for the proposed SkiM models.
In order to verify the effectiveness of Mem-LSTMs that we introduced in Sec. \ref{sec:skim_separator},
the Mem-LSTM for hidden and cell states are selectively canceled.
The corresponding initial states of the next SkiM block are replaced by $\mathbf{0}$ vector.
When both kinds of the Mem-LSTMs are canceled, the SkiM model degenerates into a naive LSTM model.
Besides replacing the states with $\mathbf{0}$ vector, we also add an experiment to replace the Mem-LSTM with an identity mapping (`id' in Table \ref{tab:ablation}) , i.e., directly pass the local states to the next SkiM block.

From the results in Table \ref{tab:ablation}, we can make at least two conclusions. 
First, both two kinds of states with global information matter in the SkiM model, and Mem-LSTMs play an important role in the SkiM model.
Second, the performance of identity mapping is almost the same with the zero initialization, and they are much worse than the original SkiM.
That means the local hidden states and cell states not processed by Mem-LSTM do not help.

\subsection{Comparison with other models on WSJ0-2mix Benchmark}

In addition to the real-time CSS task, we further test the SkiM model and compare it with other models on the utterance-level WSJ0-2mix benchmark \cite{hershey2016deep}.
We use kernel size of $2$ and $8$ in the convolutional encoders
The results are listed as `SkiM-KS2' and `SkiM-KS8' in Table \ref{tab:skim_wsj}. 
It is noted that all systems used offline models. 
Similar to the results in the CSS experiments, the SkiM models get on par performance and less computation cost than DPRNNs.
The experimental results show that SkiM is also a competitive model for utterance-level speech separation.

\begin{table}[h]
\caption{Comparison with other models on WSJ0-2mix Benchmark. (*):MACs per second estimated by us.}
\label{tab:skim_wsj}
\centering
    \setlength{\tabcolsep}{1.5mm}{
\scalebox{0.9}{
\begin{tabular}{c|cccc}
\toprule
Model & \begin{tabular}[c]{@{}c@{}} Model\\size (M)\end{tabular}  &  \multicolumn{1}{c}{\begin{tabular}[c]{@{}c@{}}MACs \\ (G/s)\end{tabular}} & SI-SNRi & SDRi   \\
\midrule
DPCL++ \cite{isik2016single} & 13.6 & - & 10.8 & -  \\
ADANet \cite{luo2018speaker} & 9.1 & - & 10.4 & 10.8 \\
WA-MISI-5 \cite{wang2018end} & 32.9 & - & 12.6 & 13.1 \\
Conv-TasNet-gLN \cite{luo2019conv} & 5.1 & 3.2$^{*}$ & 15.3 & 15.6 \\
Deep CASA \cite{liu2019divide} & 12.8 & - & 17.7 & 18.0 \\
FurcaNeXt \cite{shi2019furcanext} & 51.4 & - & - & 18.4  \\
DPRNN-KS2 \cite{luo2020dual}  & 2.6 & 38.9$^{*}$ & 18.8 & 19.0 \\
DPRNN-KS8 \cite{luo2020dual} & 2.6 & 9.8$^{*}$ & 17.0 & 17.3 \\
SepFormer \cite{subakanAttentionAllYou2021} & 26.0 & 32.1* & \textbf{20.4} & \textbf{20.5} \\
\midrule

SkiM-KS2  & 5.9 & 19.7 & 18.3 & 18.7 \\
SkiM-KS8  & 5.9 & 4.9 & 17.4 & 17.8 \\

 \bottomrule
\end{tabular}
}
}
\end{table}

\section{Conclusion}

In this paper, we explore the low-latency real-time continuous speech separation with the proposed SkiM model.
SkiM model is a simple yet effective extension on LSTM for modeling very long sequences.
The skipping memory manner for long-span information requires much less computation cost when compared to the frame-by-frame method in the DPRNN.
The experiment results show that the causal SkiM models get even better separation performance than DPRNN in the online CSS task, but with $75\%$ computational cost reduction.
Our fastest model achieves low-latency of less than $1.0$ ms on the low-power device, which shows that the proposed SkiM model is a suitable candidate for low-latency real-time CSS.

\section{Acknowledgments}
This work was supported by the National Key Research and Development Program of China (Grant No. 2021ZD0201504) and the China NSFC projects (No. 62122050 and No. 62071288).

\bibliographystyle{IEEEbib}
\bibliography{refs}

\begin{thebibliography}{10}

\bibitem{hershey2016deep}
John~R Hershey, Zhuo Chen, Jonathan Le~Roux, and Shinji Watanabe,
\newblock ``Deep clustering: Discriminative embeddings for segmentation and
  separation,''
\newblock in {\em Proc. IEEE ICASSP}, 2016, pp. 31--35.

\bibitem{yu2017permutation}
Dong Yu, Morten Kolb{\ae}k, Zheng-Hua Tan, and Jesper Jensen,
\newblock ``Permutation invariant training of deep models for
  speaker-independent multi-talker speech separation,''
\newblock in {\em Proc. IEEE ICASSP}, 2017, pp. 241--245.

\bibitem{kolbaek2017multitalker}
Morten Kolb{\ae}k, Dong Yu, Zheng-Hua Tan, and Jesper Jensen,
\newblock ``Multitalker speech separation with utterance-level permutation
  invariant training of deep recurrent neural networks,''
\newblock {\em IEEE/ACM Trans. ASLP.}, vol. 25, no. 10, pp. 1901--1913, 2017.

\bibitem{chen2017deep}
Zhuo Chen, Yi~Luo, and Nima Mesgarani,
\newblock ``Deep attractor network for single-microphone speaker separation,''
\newblock in {\em Proc. IEEE ICASSP}, 2017, pp. 246--250.

\bibitem{wang2018end}
Zhong-Qiu Wang, Jonathan~Le Roux, DeLiang Wang, et~al.,
\newblock ``End-to-end speech separation with unfolded iterative phase
  reconstruction,''
\newblock {\em arXiv preprint arXiv:1804.10204}, 2018.

\bibitem{luo2018tasnet}
Yi~Luo and Nima Mesgarani,
\newblock ``Tasnet: Time-domain audio separation network for real-time,
  single-channel speech separation,''
\newblock in {\em Proc. IEEE ICASSP}, 2018, pp. 696--700.

\bibitem{delcroix2018single}
Marc Delcroix, Katerina Zmolikova, Keisuke Kinoshita, et~al.,
\newblock ``Single channel target speaker extraction and recognition with
  speaker beam,''
\newblock in {\em Proc. IEEE ICASSP}, 2018, pp. 5554--5558.

\bibitem{luo2019conv}
Yi~Luo and Nima Mesgarani,
\newblock ``Conv-{T}as{N}et: Surpassing ideal time--frequency magnitude masking
  for speech separation,''
\newblock {\em IEEE/ACM Trans. ASLP.}, vol. 27, no. 8, pp. 1256--1266, 2019.

\bibitem{wang2019voicefilter}
Quan Wang, Hannah Muckenhirn, Kevin Wilson, et~al.,
\newblock ``{VoiceFilter: Targeted Voice Separation by Speaker-Conditioned
  Spectrogram Masking},''
\newblock in {\em Proc. Interspeech 2019}, 2019, pp. 2728--2732.

\bibitem{luo2020dual}
Yi~Luo, Zhuo Chen, and Takuya Yoshioka,
\newblock ``Dual-path {RNN}: Efficient long sequence modeling for time-domain
  single-channel speech separation,''
\newblock in {\em Proc. IEEE ICASSP}, 2020, pp. 46--50.

\bibitem{subakanAttentionAllYou2021}
Cem Subakan, Mirco Ravanelli, Samuele Cornell, et~al.,
\newblock ``Attention {{Is All You Need In Speech Separation}},''
\newblock in {\em Proc. IEEE ICASSP}, June 2021, pp. 21--25.

\bibitem{yoshioka2019css}
Takuya Yoshioka, Igor Abramovski, Cem Aksoylar, et~al.,
\newblock ``Advances in online audio-visual meeting transcription,''
\newblock in {\em Proc. IEEE ASRU}, 2019, pp. 276--283.

\bibitem{chen2020continuous}
Zhuo Chen, Takuya Yoshioka, Liang Lu, et~al.,
\newblock ``Continuous speech separation: Dataset and analysis,''
\newblock in {\em Proc. IEEE ICASSP}, 2020, pp. 7284--7288.

\bibitem{wangMultimicrophoneComplexSpectral2020}
Zhong-Qiu Wang, Peidong Wang, and DeLiang Wang,
\newblock ``Multi-microphone {{Complex Spectral Mapping}} for
  {{Utterance}}-wise and {{Continuous Speaker Separation}},''
\newblock {\em arXiv:2010.01703 [cs, eess]}, Oct. 2020.

\bibitem{li2021dualpath}
Chenda Li, Yi~Luo, Cong Han, et~al.,
\newblock ``Dual-path {RNN} for long recording speech separation,''
\newblock in {\em Proc. IEEE SLT}, 2021, pp. 865--872.

\bibitem{chenDonShootButterfly2021}
Sanyuan Chen, Yu~Wu, Zhuo Chen, et~al.,
\newblock ``Don't {{Shoot Butterfly}} with {{Rifles}}: {{Multi}}-{{Channel
  Continuous Speech Separation}} with {{Early Exit Transformer}},''
\newblock in {\em Proc. IEEE ICASSP}, June 2021, pp. 6139--6143.

\bibitem{liDualPathModelingLong2021}
Chenda Li, Zhuo Chen, Yi~Luo, et~al.,
\newblock ``Dual-{{Path Modeling}} for {{Long Recording Speech Separation}} in
  {{Meetings}},''
\newblock in {\em Proc. IEEE ICASSP}, June 2021, pp. 5739--5743.

\bibitem{chenContinuousSpeechSeparation2021}
Sanyuan Chen, Yu~Wu, Zhuo Chen, et~al.,
\newblock ``Continuous {{Speech Separation}} with {{Conformer}},''
\newblock in {\em Proc. IEEE ICASSP}, June 2021, pp. 5749--5753.

\bibitem{chenUltraFastSpeech2021}
Sanyuan Chen, Yu~Wu, Zhuo Chen, et~al.,
\newblock ``Ultra {{Fast Speech Separation Model}} with {{Teacher Student
  Learning}},''
\newblock in {\em Interspeech 2021}. Aug. 2021, pp. 3026--3030, {ISCA}.

\bibitem{hanContinuousSpeechSeparation}
Cong Han, Yi~Luo, Chenda Li, et~al.,
\newblock ``Continuous speech separation using speaker inventory for long
  recording,''
\newblock in {\em Proc. ISCA Interspeech}, 2021, p.~5.

\bibitem{neumann21_interspeech}
Thilo von Neumann, Keisuke Kinoshita, Christoph Boeddeker, et~al.,
\newblock ``{Graph-PIT: Generalized Permutation Invariant Training for
  Continuous Separation of Arbitrary Numbers of Speakers},''
\newblock in {\em Proc. ISCA Interspeech}, 2021, pp. 3490--3494.

\bibitem{vaswani2017attention}
Ashish Vaswani, Noam Shazeer, Niki Parmar, et~al.,
\newblock ``Attention is all you need,''
\newblock in {\em Advances in neural information processing systems}, 2017, pp.
  5998--6008.

\bibitem{hochreiter1997long}
Sepp Hochreiter and J{\"u}rgen Schmidhuber,
\newblock ``Long short-term memory,''
\newblock {\em Neural computation}, vol. 9, no. 8, pp. 1735--1780, 1997.

\bibitem{he2016identity}
Kaiming He, Xiangyu Zhang, Shaoqing Ren, and Jian Sun,
\newblock ``Identity mappings in deep residual networks,''
\newblock in {\em European Conference on Computer Vision}. Springer, 2016, pp.
  630--645.

\bibitem{panayotov2015librispeech}
Vassil Panayotov, Guoguo Chen, Daniel Povey, et~al.,
\newblock ``Librispeech: an {ASR} corpus based on public domain audio books,''
\newblock in {\em Proc. IEEE ICASSP}. IEEE, 2015, pp. 5206--5210.

\bibitem{wisdom2020unsupervised}
Scott Wisdom, Efthymios Tzinis, Hakan Erdogan, Ron~J Weiss, Kevin Wilson, and
  John~R Hershey,
\newblock ``Unsupervised sound separation using mixtures of mixtures,''
\newblock {\em arXiv preprint arXiv:2006.12701}, 2020.

\bibitem{li2021espnet}
Chenda Li, Jing Shi, Wangyou Zhang, et~al.,
\newblock ``{ESPNet-SE}: End-to-end speech enhancement and separation toolkit
  designed for {ASR} integration,''
\newblock in {\em Proc. IEEE SLT}, 2021, pp. 785--792.

\bibitem{kingma2014adam}
Diederik Kingma and Jimmy Ba,
\newblock ``Adam: A method for stochastic optimization,''
\newblock {\em arXiv preprint arXiv:1412.6980}, 2014.

\bibitem{taal2010short}
Cees~H Taal, Richard~C Hendriks, Richard Heusdens, et~al.,
\newblock ``A short-time objective intelligibility measure for time-frequency
  weighted noisy speech,''
\newblock in {\em Proc. IEEE ICASSP}. IEEE, 2010, pp. 4214--4217.

\bibitem{isik2016single}
Yusuf Isik, Jonathan Le~Roux, Zhuo Chen, Shinji Watanabe, and John~R Hershey,
\newblock ``Single-channel multi-speaker separation using deep clustering,''
\newblock {\em Proc. ISCA Interspeech}, pp. 545--549, 2016.

\bibitem{luo2018speaker}
Yi~Luo, Zhuo Chen, and Nima Mesgarani,
\newblock ``Speaker-independent speech separation with deep attractor
  network,''
\newblock {\em IEEE/ACM Trans. ASLP.}, vol. 26, no. 4, pp. 787--796, 2018.

\bibitem{liu2019divide}
Yuzhou Liu and DeLiang Wang,
\newblock ``Divide and conquer: A deep {CASA} approach to talker-independent
  monaural speaker separation,''
\newblock {\em IEEE/ACM Trans. ASLP.}, vol. 27, no. 12, pp. 2092--2102, 2019.

\bibitem{shi2019furcanext}
Liwen Zhang, Ziqiang Shi, Jiqing Han, et~al.,
\newblock ``{FurcaNeXt}: End-to-end monaural speech separation with dynamic
  gated dilated temporal convolutional networks,''
\newblock in {\em MultiMedia Modeling}. 2020, vol. 11961, pp. 653--665,
  Springer.

\end{thebibliography}

\end{document}